\documentclass[aps,prb,twocolumn, showpacs,superscriptaddress,citeauthorscript]{revtex4-2}

\usepackage{graphicx} 
\usepackage{dcolumn} 
\usepackage{bm}        
\usepackage{amssymb}   
\usepackage{amsmath}
\usepackage{gensymb}
\usepackage{color}  
\usepackage[normalem]{ulem}
\usepackage[T1]{fontenc}
\renewcommand{\vec}[1]{\boldsymbol{\mathbf{#1}}}

\newcommand{\veca}[1]{\overrightarrow{#1}}
\usepackage{hyperref}

\usepackage[dvipsnames]{xcolor}
\setcitestyle{super,open={},close={}}
\usepackage[normalem]{ulem}

\hypersetup{
    colorlinks,
    linkcolor={blue!50!black},
    citecolor={blue!50!black},
    urlcolor={blue!80!black}
}

\begin{document}

\title{A hole spin resilient to dipole-induced thermal effects}

\author{V. Champain}\thanks{Now at: ICFO - Institut de Ciencies Fotoniques, The Barcelona Institute of Science and Technology, 08860 Castelldefels (Barcelona), Spain.}
\affiliation{Univ. Grenoble Alpes, CEA, Grenoble INP, IRIG-Pheliqs, Grenoble, France.}
\author{G. Boschetto}
\author{H. Niebojewski}
\author{B. Bertrand}
\affiliation{Univ. Grenoble Alpes, CEA, Leti, F-38000 Grenoble, France.}
\author{L. Mauro}
\affiliation{Univ. Grenoble Alpes, CEA, IRIG-MEM-L\textunderscore Sim, Grenoble, France.}
\author{C. Haab}
\author{M. Bassi}\thanks{Now at: QuTech and Kavli Institute of Nanoscience, Delft University of Technology, 2600 GA Delft, The Netherlands}
\author{V. Schmitt}
\author{X. Jehl}
\author{S. Zihlmann}
\author{R. Maurand}
\affiliation{Univ. Grenoble Alpes, CEA, Grenoble INP, IRIG-Pheliqs, Grenoble, France.}
\author{Y.-M. Niquet}
\affiliation{Univ. Grenoble Alpes, CEA, IRIG-MEM-L\textunderscore Sim, Grenoble, France.}
\author{C. B. Winkelmann}
\author{S. De Franceschi}
\affiliation{Univ. Grenoble Alpes, CEA, Grenoble INP, IRIG-Pheliqs, Grenoble, France.}
\author{B. Martinez}
\affiliation{Univ. Grenoble Alpes, CEA, Leti, F-38000 Grenoble, France.}
\author{B. Brun}
\email{E-mail: boris.brun-barriere@cea.fr}
\affiliation{Univ. Grenoble Alpes, CEA, Grenoble INP, IRIG-Pheliqs, Grenoble, France.}

\begin{abstract}
Recent advances in scaling up spin-based quantum processors have revealed unanticipated issues related to thermal effects. Microwave pulses required to manipulate and read the qubits are found to overheat the spins environment, which unexpectedly induces Larmor frequency shifts, reducing thereby gate fidelities.
In this study, we shine light on these elusive thermal effects, by experimentally characterizing the temperature dependence of the Larmor frequency for a single hole spin in silicon. Our results unambiguously reveal an electrical origin underlying the thermal susceptibility, stemming from the spin-orbit-induced electric susceptibility. 
We perform an accurate modeling of the spin electrostatic environment and gyromagnetic properties, allowing us to pinpoint electric dipoles which unfreeze as the temperature increases, as responsible for these frequency shifts. 
Surprisingly, we find that the thermal susceptibility can be tuned with the magnetic field angle and can even cancel out, unveiling a sweet spot where the hole spin is rendered immune to thermal effects. 
These findings reveal the microscopic origin of longitudinal thermal susceptibility, providing key insight for the design of thermally robust semiconductor qubits.
\end{abstract}

\maketitle

\section{Introduction}

Spins in semiconducting materials offer promising prospects for large-scale quantum computing due to their compact size and compatibility with standard industrial manufacturing processes. Recently, group-IV-based quantum dots have taken a step further with the demonstration of functional spin-qubit arrays with up to six\cite{philips_universal_2022} and ten\cite{john_robust_2025} qubits in silicon and germanium quantum wells, respectively. In parallel, single- and two-qubit gate fidelities have largely overcome the fault-tolerance threshold for quantum error correction \cite{yoneda_quantum-dot_2018, veldhorst_silicon_2017, xue_quantum_2022, noiri_fast_2022,mills_two-qubit_2022}.  This significant progress makes semiconductor spin qubits a credible candidate in the endeavor for scalable quantum computing \cite{vandersypen_interfacing_2017}. The simultaneous operation of several spin qubits,  however, has revealed unanticipated issues. It has been shown that microwave as well as baseband control pulses cause unintentional local heating \cite{champain_real-time_2024,ye_measuring_2026}, accompanied by significant variations in the qubit Larmor frequencies. This leads to  dephasing and a consequent loss of gate fidelities in simultaneously operated spin qubits \cite{freer_single_2017,philips_universal_2022, zwerver_qubits_2022,  lawrie_simultaneous_2023, savytskyy_electrically_2023}, pinpointing a clear hurdle for spin-qubit integration.

Recently, Undseth et al. systematically investigated the temperature dependence of the Larmor frequency in six neighboring electron spins in silicon \cite{undseth_hotter_2023}, electrically driven by means of the synthetic spin-orbit coupling induced by a micromagnet. All Larmor frequencies were found to exhibit a nonmonotonic temperature dependence, increasing with temperature in the range from 20 mK to roughly 200 mK, and decreasing at higher temperatures. This results in a temperature sweet spot around 200 mK where the Larmor frequency is first-order temperature independent rendering qubit performance robust against local heating. 

To date, the physical mechanism underlying the observed temperature dependence of the Larmor frequency remains unclear. Recent theoretical works \cite{choi_interacting_2024, sato_simulation_2024} have proposed a phenomenological model based on thermally activated electric dipoles. By creating a temperature-dependent electric field, such dipoles alter electrical confinement, causing  a small, yet relevant displacement of the electron wavefunction in the magnetic field gradient generated by the micromagnet. It is shown that the resulting variation in the effective magnetic field experienced by the electron can account for the observed temperature dependence of its Larmor frequency. 

In this work, we investigate the thermal susceptibility of a single hole spin in a silicon metal-oxide-semiconductor (MOS) nanowire device. Owing to their intrinsic spin-orbit coupling, hole spins exhibit a strongly anisotropic and gate-dependent gyromagnetic factor \cite{voisin_few-electron_2014,bogan_consequences_2017,liles_electrical_2021}. As a consequence, the Larmor frequency generally varies with gate voltage. This dependence is quantified by the gate-voltage derivative of the Larmor frequency, known as the longitudinal spin-electric susceptibility (LSES). It was recently found that the LSES can vanish for specific magnetic field directions, defining operational sweet spots where hole spins are first-order-insensitive to charge noise\cite{piot_single_2022, bassi_optimal_2026}. In a similar way, here we study the temperature dependence of the hole Larmor frequency for different magnetic field directions. 

We find that the Larmor thermal susceptibility (LTS), defined as the temperature derivative of the Larmor frequency, can be as large as $\sim10~\rm MHz/K$, i.e. the same order of magnitude as the one observed for electron spins in silicon quantum wells\cite{undseth_hotter_2023}. Interestingly, 
upon varying the magnetic field orientation we observe a clear correlation between electrical and thermal susceptibilities, unveiling the existence of a common electrical origin mediated by the spin-orbit interaction. Along the same line as Ref.\citep{choi_interacting_2024}, we propose a microscopic model involving a surrounding dipole bath that unfreezes as temperature is increased, changing the electrostatic environment of the hole spin qubit. By numerically calculating the full three-dimensional electrostatics of the device including a random distribution of these dipoles we achieve a good agreement with the experimental data, and find a surprisingly small dipole amplitude of the order of 1 $e\cdot\rm  pm$. We also reveal the existence of a "thermal sweet spot", a particular magnetic field orientation for which the Larmor frequency does not depend on temperature.\\

\section{Experimental results}\label{sec:method}

\begin{figure}
    \centering
    \includegraphics{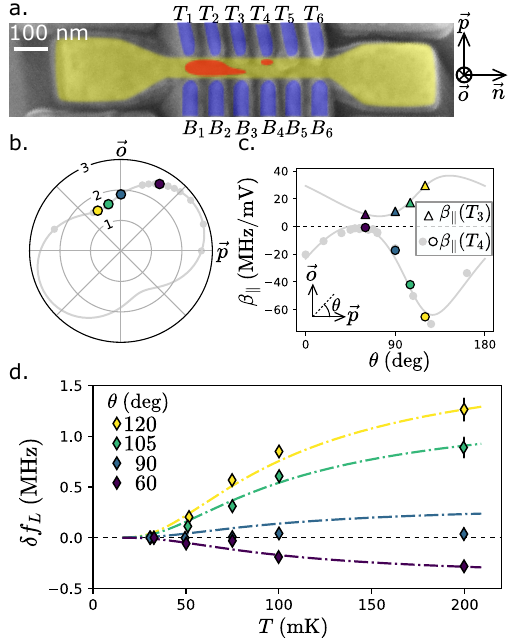}
    \caption{\textbf{Device characterization and temperature dependence. a.} Top view of the device. Blue: Accumulation gates. Yellow: Silicon nanowire, the narrow part is undoped and fully depleted while the sides are highly positively doped. Red: figurative hole accumulation in the charge sensor and in the dot hosting a single hole. \textbf{b.} (resp. \textbf{c.}) g-factor (resp. longitudinal spin electric susceptibilities) as a function of the magnetic field orientation. Scattered points are measured data, and the solid lines are fits (see Appendix~\ref{app:fit} for the fit formulas). The colored circles and triangles correspond to the selected orientations of the field for which we measure the temperature dependence of the Larmor frequency. \textbf{d.} Larmor frequency shift measured as a function of temperature, defined as $ \delta f_L = f_L(T) - f_L(30~\rm mK)$. The Larmor frequency is extracted by averaging Ramsey oscillations over 30 minutes and the inhomogeneous dephasing times extracted from the same measurement are analyzed in Appendix~\ref{app:T2}. The dot-dashed line correspond to the fits discussed in section~\ref{sec:electric_fit}. The dashed black line shows zero Larmor shift. }
    \label{fig1}
\end{figure}

Our device consists of an undoped silicon-on-insulator nanowire with a $17 \times 40$ nm$^2$ cross-section, terminated on both edges by degenerately doped p-type regions, providing the necessary hole reservoirs. The electrostatic potential in the silicon channel is controlled by means of six pairs of face-to-face gates labeled as $\{T_i,B_i\}$ with $i=1,2,...,6$ (see Fig.~\ref{fig1}a).  RF reflectometry charge readout is performed using a resonator galvanically connected to the left lead. A single hole is trapped in a quantum dot (QD) next to gate $T_4$. Its spin state is measured via spin-selective tunneling \cite{elzerman_single-shot_2004} towards a relatively large single-hole box (SHB) defined by gates $T_1$, $T_2$, $B_1$, $B_2$ and $B_3$ (see Fig.~\ref{fig1}a). This readout method limits the accessible temperature range to 200 mK.
This SHB acts both as a hole reservoir for the QD and as a charge sensor, its electrochemical potential being aligned with the Fermi energy of the left  lead.   
The device is thermally anchored to the mixing chamber of a dilution refrigerator with a base temperature of 30 mK. The cryostat is equipped with a two-axis vector magnet that allows varying the magnetic field angle, $\theta$, in the ($\veca{o}$,$~\veca{p}$) plane perpendicular to the nanowire axis $\veca{n}$ (the geometrical axes $~\veca{o},~\veca{p},~\veca{n}$ and $\theta$ are precisely defined in Figs.~\ref{fig1}a and ~\ref{fig1}c). For each orientation, the magnetic field amplitude is adjusted to maintain a constant Larmor frequency $f_L = 17.1$ GHz.

We begin by characterizing the hole g-factor anisotropy in the ($\veca{o},\veca{p}$) plane. The results, shown in Fig.~\ref{fig1}b, reveal a g-factor varying between about 1.5 and 2.9, with principal axes titled with respect to the device geometric axes (most likely due to shear strain), as already observed in earlier experiments \cite{piot_single_2022, bassi_optimal_2026}. Following a procedure described in Ref. \cite{piot_single_2022}, we then measure the anisotropy of the LSES associated with variations in the voltage on a gate $G_i$

\begin{equation}
    \beta_\parallel (G_i) = \frac{\partial f_L}{\partial V_{G_i}}.
\end{equation}
We consider here voltage variations on gates $T_4$ and $T_3$, quantifying the sensitivity of $f_L$ to variations in the electric field perpendicular to, and along the nanowire axis, respectively. The LSESs $\beta_\parallel(T_4)$ and $\beta_\parallel (T_3)$ are shown in Fig.~\ref{fig1}c and respectively range from $-70$ to $0~\rm MHz/mV$ and from $10$ to $40~\rm MHz/mV$. 

We select four distinct magnetic field orientations where $\beta_\parallel(T_4)$ takes well-separated values, as indicated by the colored symbols in Fig.~\ref{fig1}b and c. For each of these orientations, we measure the temperature dependence of $f_L$ by repeating a Ramsey interferometry protocol (similar to Ref.\cite{undseth_hotter_2023}) at various temperatures of the cryostat mixing chamber. Following each temperature change, the system is allowed to thermalize for 30 minutes. Notably, the gate voltages are kept constant throughout the full range of explored temperatures. The measured variation of the Larmor frequency relative to the base-temperature value, $\delta f_L$, is plotted in Fig.~\ref{fig1}d for the selected magnetic field angles. This data set is the main experimental outcome of the present work.

The first important observation is that the hole Larmor frequency exhibits a clear dependence on temperature, i.e. a generally finite LTS. The largest variation of the Larmor frequency ($\delta f_L \sim 1.25$ MHz) occurs at $\theta = 120~\rm deg$ (yellow diamonds in Fig.~\ref{fig1}d), with a LTS as large as 12 MHz/K. This thermal susceptibility is comparable to the largest LTS values reported in Ref.~\cite{undseth_hotter_2023} for electrons in silicon quantum wells, despite major differences in the two physical systems. To the best of our knowledge, the observation of a finite LTS for hole spins has not been reported before.

The second observation is that the LTS clearly depends on the magnetic field orientation. This behavior is generally similar to the one observed for the LSES. With a closer look, we can notice a correlation in the angular dependence of the two susceptibilities. Both the LTS and $\beta_\parallel(T_4)$ reach their largest amplitude around  $\theta = 120~\rm deg$ and are much smaller close to $\theta = 60~\rm deg$. This correlation suggests that the observed LTS originates from temperature-dependent local electric fields acting on the hole spin via spin-orbit coupling.

Interestingly, Fig.~\ref{fig1}d shows that the Larmor frequency shift vanishes at $\theta = 90~\rm deg$ (blue diamonds). This magnetic field orientation defines a thermal sweet spot where the hole Larmor frequency is temperature independent all over the explored temperature range. The observed thermal sweet spot provides an optimal setting for heat-resilient operation, as further discussed in Sec. \ref{sec:discussion}.

The inferred electrical origin of the experimentally observed LTS is compatible with the physical picture of an environment of two-level systems acting as thermally activated electric dipoles proposed in Refs.\cite{choi_interacting_2024, sato_simulation_2024}. In the following, building upon this basic idea, we shall first elucidate the link between LTS and LSES on a phenomenological level. Then, using realistic numerical simulations, we shall demonstrate that an ensemble of electric dipoles in the device dielectric layer can remarkably well account for the observed LTS as well as for the existence of thermal sweet spots.

\section{Random-Field Dipole Bath}\label{sec:model}

\subsection{Phenomenological evidence} \label{sec:electric_fit}

An ensemble of dipoles generates an electric field whose average magnitude depends on the relative occupation of their two accessible states, which follows a Boltzmann distribution. Let us assume that the effect of these dipoles can be cast into effective changes in the voltages on the different gates:

\begin{equation}
    \delta V_{G_i} =  \delta V_{0,G_i} \left(1 - \tanh\left(\frac{T_{0,G_i}}{T}\right)\right),
    \label{eq:boltz}
\end{equation}
where $T_{0,G_i}$ is the average activation temperature associated with the dipoles near gate $G_i$, and $\delta V_{0,G_i}$ is the voltage variation  emulating the transition from a fully polarized to a fully depolarized dipole ensemble. With this assumption, the effect of such effective gate-voltage variations on the Larmor frequency is directly proportional to the LSES. For the hole spin qubit studied here, we only consider voltage variations on gates $T_3$ and $T_4$, used as proxies to emulate electric-field variations perpendicular and longitudinal to the nanowire axis, respectively.   

In a first attempt to understand the physical mechanisms at play with a minimal set of fitting parameters, we consider identical effective gate voltage shifts, i.e. $\delta V_{T_3} = \delta V_{T_4} = \delta V_0$, and that there is a single activation temperature $T_0$ for all dipoles. Following these rough approximations, we can fit the measured shifts in the Larmor frequency using the simple relation 

\begin{equation}
    \delta f_L(T) = \delta V_0 \left( \beta_{\parallel}(T_3) + \beta_{\parallel}(T_4) \right) \left( 1 - \tanh\left( \frac{T_0}{T} \right) \right), \label{eq:freqshift_model}
\end{equation}
where $\beta_{\parallel}(T_3)$ and $\beta_{\parallel}(T_4)$ are measured  quantities, and $\delta V_0$ and $T_0$ are the only two fitting parameters.
The results of this fit are shown as dot-dashed lines in Fig.~\ref{fig1}d. They show that our simple phenomenological model can reproduce fairly well the observed angular dependence of the thermally-induced shift in the qubit Larmor frequency. 
The fit parameters are $\delta V_0 = 55 \pm 4~\mu\mathrm{V}$ and $T_0 = 70 \pm 5~\mathrm{mK}$. While the agreement is very good, we note that additional mechanisms, such as Overhauser fields (see Appendix \ref{app:hyperfin}), may also contribute at a smaller scale.
Interestingly,  Eq.~\eqref{eq:freqshift_model} can also account for the existence of a thermal sweet spot, which is expected for specific magnetic field orientations where $\beta_{\parallel}(T_3) = -\beta_{\parallel}(T_4)$. 

To go beyond this phenomenological model and gain quantitative insight into the dipole bath, we now turn to numerical simulations on a realistic model of the device structure. To this aim, we use a a three-dimensional solver of Poisson and Schrödinger equations based on a six-band $\vec{k} \cdot \vec{p}$ Hamiltonian, already introduced in Refs. ~\cite{venitucci_electrical_2018, piot_single_2022}.

\subsection{Simulation}

\begin{figure}
    \centering
    \includegraphics{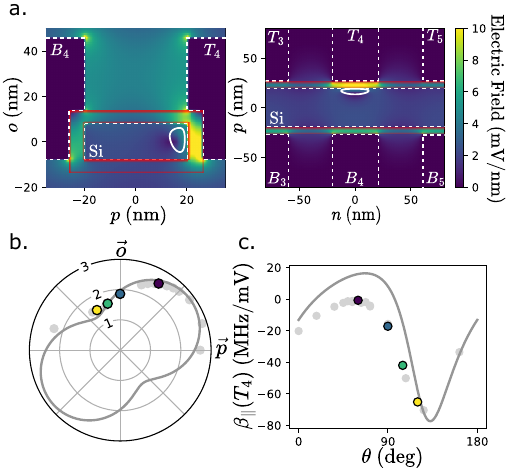}
    \caption{\textbf{Simulation of the device. a.} Electric field map of the device. The white dashed lines delimitate the gates and the silicon channel, in between are the oxides and dielectrics. The solid white line depicts the hole wavefunction and the faded red lines delimit the 6-nm-thick shell where dipoles are placed. The colorbar is shared for both panels. \textbf{b.} (resp. \textbf{c.}) g-factor (resp. longitudinal spin electric susceptibility) as a function of the magnetic field orientation. The scattered points correspond to the experimental data, and the solid grey lines are the simulation results.} 
    \label{fig2}
\end{figure}

As a first step,  we need to set the device parameters in order to reproduce the experimentally observed charge and spin characteristics of the hole quantum dot. A single hole quantum dot is simulated at the MOS interface with $T_4$ by tuning the gate voltages on gates $T_3-T_5$ and $B_3-B_5$, see Fig.~\ref{fig2}a. Then, electrostatic confinement and shear strains are further adjusted in such a way that the simulated $g$-factor anisotropy matches reasonably well the measured one (see Fig.~\ref{fig2}b). With these parameters fixed, we compute $\beta_\parallel(T_4)$ without any additional tuning. The simulated angular dependence, plotted with a solid line Fig.~\ref{fig2}c, approaches reasonably well the experimental data. The quantitative differences (a clear sign change and two sweet spots $\beta_\parallel(T_4)=0$) may be due to a misalignment along the third magnetic axis, not explored experimentally. The calculation of $\beta_\parallel(T_3)$ and its comparison to the experiment are discussed in Appendix~\ref{app:LSEST3}.

Overall, the results of our simulations, when considered in their ensemble (electrostatics, g-factors, and LSES), are in reasonably good agreement with the experimental findings, highlighting a faithful modeling of the electric fields in the vicinity of the hole quantum dot.

We now turn to the simulation of dipole-induced effects on the hole spin. Following a methodology similar to that of previous studies~\cite{choi_interacting_2024, sato_simulation_2024}, we consider a statistical ensemble of non-interacting dipoles, which respond to the electric field created by the gates and by the charge confined in the quantum dot. We only consider dipoles within a 6-nm-thick domain around the nanowire (the front gates oxide thickness) where the electric field is the strongest. The positions and orientations of these dipoles are uniformly sampled on a grid within the oxide. We place a single dipole on a grid site and solve the six-band $\vec{k} \cdot \vec{p}$ Hamiltonian numerically, to extract its impact on the Larmor frequency. We repeat this procedure for each site and dipole orientation and compare the obtained Larmor frequency with the bare one, i.e without the dipole. This yields an ensemble of frequency shifts, $\Delta f_{L,i}$, where the index $i$ labels the dipoles.

We can then generate a distribution of $N$ dipoles located at random positions within the oxide, which yields to a temperature-dependent total frequency shift: 
\begin{equation}
\delta f_{L}^{(N)} (T)= \sum_{i=1}^{N} -\Delta f_{L,i} \tanh\left(\frac{\varepsilon_i}{k_B T}\right), \label{eq:temp_dep}
\end{equation}
where $\varepsilon_i$ is the energy of the $i$-th dipole. 
This energy is given by
\begin{equation}
\varepsilon_i =  \varepsilon_{0,i} - \vec{p}_i \cdot \vec{E}_i  , \label{eq:def_energy}
\end{equation}
with $\vec{p}_i$ the dipole moment, $\vec{E}_i$ the electric field at the $i$-th dipole position, and $\varepsilon_{0,i}$ a possible intrinsic detuning of the dipole in absence of electric field. Note that the impact of each dipole on the Larmor frequency is finite for $|\varepsilon_i| \gg k_B T$ and vanishes when $|\varepsilon_i| \ll k_B T$.

 At this stage, we introduce two simplifying assumptions. First, we assume that $ |\varepsilon_{0,i}| \ll |\vec{p} \cdot \vec{E}_i|$ so that Eq.~\eqref{eq:def_energy} becomes $\varepsilon_i \approx  -\vec{p}_i \cdot \vec{E}_i$. 
Importantly, since an intrinsic detuning $\varepsilon_{0,i}$ doesn't set a preferential direction, a finite LTS requires $\varepsilon_i$ to be dominated by the $\vec{p}_i \cdot \vec{E}_i$.

 Second, we assume for simplicity that all dipoles have the same moment $|\vec{p}_i|=|\vec{p}|$. 
We have verified that these assumptions do not qualitatively affect our conclusions, as discussed in Appendix \ref{app:model_hom}, \ref{app:model} and \ref{app:d}. 

\begin{figure}
    \centering
    \includegraphics[width=\linewidth]{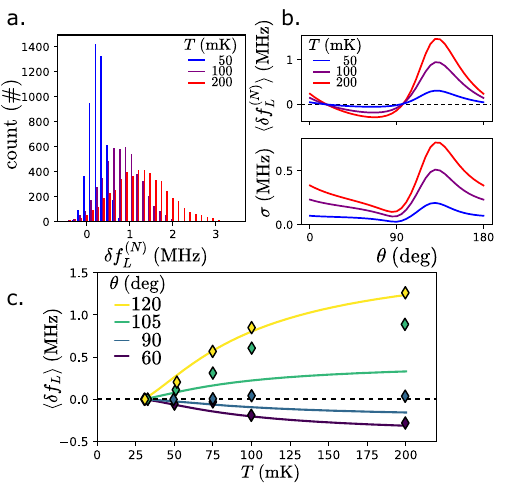}
    \caption{\textbf{Dipole model simulation. a.}  Histograms of $\delta f_L^{(N)}(T)$ normalized with respect to $\delta f_L^{(N)}(30~$mK$)$ for $5000$ different dipoles configurations with $N=460$ dipoles ($\rho = 3.54 \times 10^{18}$~cm$^{-3}$) and $|\vec{p}| = 0.6~e\cdot$pm, at 50 mK (blue), 100 mK (purple) and 200 mK (red). The magnetic field orientation is fixed at $\theta=120~\rm deg$ .  \textbf{b.} Dependence on $\theta$ of the average $\langle\delta f_L^{(N)}\rangle$ (top panel) and of the standard deviation $\sigma$ (bottom panel). The dashed black line shows zero Larmor shift. \textbf{c.} Simulated average Larmor frequency shift as a function of temperature for the four orientations highlighted by vertical colored dashed lines in panel \textbf{b}. The colored diamonds correspond to the experimental data shown in Fig.~\ref{fig1}.}
    \label{fig3}
\end{figure}

We then generate different configurations of $N$ randomly distributed dipoles, which provide a statistical ensemble of nominally identical, yet microscopically distinct devices. For each device, we compute the total Larmor frequency shift $\delta f_L^{(N)}(T)$ as a function of temperature and magnetic field orientation, taking the reference temperature as $T = 30~\mathrm{mK}$ for consistency with the experimental data. Fig.~\ref{fig3}a shows histograms of $\delta f_L^{(N)}$ calculated on a statistical ensemble of 5000 devices for three different temperatures (50 mK,  100 mK, and 200 mK) chosen within the experimentally explored range, and $\theta = 120~\rm deg$.

As a general trend, we find scattered $\delta f_L^{(N)}$ distributions with fairly large standard deviation  $\sigma$,  centered around a temperature-dependent average value $\langle \delta f_L ^{(N)}\rangle$. Both  $\langle \delta f_L ^{(N)}\rangle$ and $\sigma$ exhibit a significant dependence on the magnetic field orientation, as shown in the upper and lower panel of Fig.~\ref{fig3}b, respectively. In each panel we show  three curves calculated for $T= 50$, 100, and 200 mK.

A clear correlation is observed between the computed $\langle \delta f_L ^{(N)}\rangle$ and the longitudinal spin-electric susceptibility $\beta_\parallel(T_4)$ (see Fig.~\ref{fig2}c and ~\ref{fig3}b), in agreement with the phenomenological model discussed above and with the experimental trend. The finite LTS is a consequence of the structural order of the dipoles aligning with the direction of the electric field at low enough temperatures, which results in an enhancement of $\vec{E}$ in the QD. Indeed, these dipoles do screen the field in the oxide, but enhance it outside. This effect therefore maps very well to effective gate voltage variations, which explains the clear correlations with the LSES and highlights the relevance of the phenomenological model in Section \ref{sec:electric_fit}.
Remarkably, the numerical results  reproduce the observed change of sign in the thermal susceptibility, which results in two thermal sweet spots ($\langle \delta f_L ^{(N)}\rangle = 0$). Moreover, the statistical variations of $\delta f_L$ due to different dipole baths turn out to be minimal at one sweet spot, as revealed by the reduced $\sigma$. It is important to note, however, that the calculated $\sigma$ doesn't account for potential qubit-to-qubit variability, which may indeed provide distinct LSES for each qubit and consequently further scatter the thermal sweet-spots \cite{martinez_variability_2022, bassi_optimal_2026}.

In Fig.~\ref{fig3}c, we compare the calculated temperature dependence of the Larmor-frequency shift to the experimental data reproduced from Fig.~\ref{fig1}d. We adjust the dipole moment, $|\vec{p}|$, and the number of dipoles, $N$, in order to reproduce the experimental data. Note that for a given electric field, the activation temperature $T_0$ depends solely on $|\vec{p}|$. We therefore first tune this parameter to achieve $T_0 \approx 70$~mK, and subsequently adjust the number of dipoles $N$ to match the experimental amplitude of $\delta f_L$. An overall good agreement is obtained for $|\vec{p}| = 0.6~e\cdot$pm and a dipole density $\rho = 3.54 \times 10^{18}$~cm$^{-3}$ (corresponding to a areal density $\sigma \sim 2.12 \times 10^{12} \,$cm$^{-2}$, or to $N=460$ dipoles randomly distributed within the 6-nm-thick oxide embedding the Si channel), as shown by solid lines in Fig.~\ref{fig3}c. We attribute the observed quantitative differences to discrepancies between the electrical susceptibilities in experiment and simulation (see Fig. 2c and Appendix D).

The inferred dipole moment $|\vec{p}|$ turns out to be surprisingly small. For a charge $e$,  this moment would correspond to a charge displacement of barely a pm. The physical origin of such small dipoles remains an open question. Their magnitude appears incompatible with typical hopping events between neighboring charge defects. They could, alternatively, result from small atomic reconfigurations within the complex potential landscape of charged defects and disordered ionic oxide materials (such as slight changes  in chemical bonds lengths or angles), or from local excitations of the dielectric polarization. Nevertheless, the distribution of such dipoles is unlikely to be limited to very low energies such that $|\varepsilon_{0,i}| \ll |\vec{p}_i \cdot \vec{E}_i|$ (see Appendix \ref{app:model} for a detailed discussion). Additionally, this distribution would most likely exhibit correlations between $|\vec{p}_i|$ and $\varepsilon_{0,i}$ not included in the present model. The physical origin of these dipoles therefore calls for further investigation.

\section{Mitigating thermal dissipation} \label{sec:discussion}
Thermal susceptibility renders the Larmor frequency sensitive to temperature variations. Therefore, the heat generated by control signals hampers gate fidelities \cite{lawrie_simultaneous_2023, sato_simulation_2024, wu_simultaneous_2025}.
To quantify this effect, we perform randomized benchmarking simulations in Appendix~\ref{app:RB}. These simulations indicate that for typical Rabi frequencies of a few MHz, thermally induced frequency shifts comparable to those reported here and in Ref.~\cite{undseth_hotter_2023} can limit fidelity to 99 \%, clearly illustrating the performance degradation caused by thermally-induced detuning and the consequent need to mitigate this effect.

In this section, we briefly review different mitigation strategies and their impact on qubit operation. We first discuss control-based approaches, such as pulse engineering, which aim to compensate for thermally induced errors at the level of qubit manipulation. We then introduce the concept of thermal sweet spots, where the qubit becomes intrinsically insensitive to temperature variations.

\subsection{Pulse engineering}

 Previous studies have proposed to saturate the environment by using off-resonant pre-pulses\cite{philips_universal_2022,kawakami_gate_2016,takeda_optimized_2018, wu_simultaneous_2025} to prepare a heated steady state where the Larmor frequency could be calibrated. Heating depends on \emph{where} the pulses are applied and \emph{when}, since quantum algorithms comprise idling times. In general, the steady state is therefore qubit- and sequence-dependent. Another approach consists in operating devices at elevated temperatures around 200 mK to reduce dissipation-related dephasing, achieving improved single-qubit gate fidelity\cite{undseth_hotter_2023}. However, this increase in temperature usually comes at the cost of reduced coherence times\cite{yang_operation_2020, camenzind_hole_2022, huang_high-fidelity_2024}.
 
 Importantly, a distinction must be made between coherent and incoherent thermal effects. 
If reproducible and quasi-static, thermally-induced frequency shifts lead to coherent errors that can be compensated using adaptive calibration, pulse shaping, or real-time phase updates \cite{takeda_optimized_2018}. 
In contrast, stochastic temperature fluctuations occurring on timescales comparable to or shorter than gate duration produce incoherent dephasing, which cannot be fully corrected by reshaping the control pulses. While control-based mitigation can partially alleviate coherent LTS-induced errors, reducing the underlying thermal susceptibility at the device level remains a compelling route toward robust high-fidelity operation.

\subsection{Thermal sweet-spot}

Blue diamonds in Fig.~\ref{fig1}d feature the most important result of our work, by revealing a specific magnetic field orientation where the Larmor frequency becomes temperature-independent, which we have coined thermal sweet-spot. This is further supported by the simulation in Fig.~\ref{fig2}b. In this peculiar configuration, the hole spin is insensitive to the average static polarization of the dipole environment.  Such a cancellation of thermal susceptibility could enable suppression of heat-induced gate errors while still operating the device at the lowest possible temperature, thereby possibly maximizing gate fidelities. 

\subsection{Matching electrical and thermal sweet-spots}

\begin{figure}
    \centering
    \includegraphics{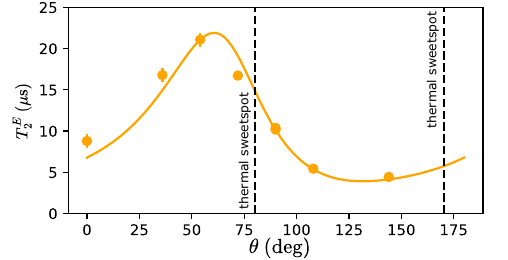}
    \caption{Hahn Echo coherence times as a function of the magnetic field orientation. The solid line represent a fit to Eq.~\eqref{eq:t2echo}. The vertical dashed black lines represent the position of the thermal sweetspots expected from the measured values of $\beta_\parallel(T_3)$ and $\beta_\parallel(T_4)$.}
    \label{fig4}
\end{figure}

Nevertheless, this thermal sweet-spot is not aligned with a coherence sweet-spot where the hole spin is largely insensitive to charge noise \cite{piot_single_2022, hendrickx_sweet-spot_2024, bassi_optimal_2026}, and we discuss below this discrepancy. 

Fig. \ref{fig4} displays the Hahn-Echo coherence time as a function of magnetic field orientation, revealing values ranging from 5 $\mu$s to 20 $\mu$s, with a maximum around $\theta = 50~\rm deg$. 
Under the assumption of linear coupling to $1/f$ charge noise, this echo coherence time can be fully captured by the LSES and writes\cite{ithier_decoherence_2005}:
\begin{equation}
    T_2^E \propto \left (\sum _{G_i}\beta_\parallel(G_i)^2S_{G_i}\right )^{-\frac{1}{2}} ,\label{eq:t2echo}
\end{equation}
where $S_{G_i}$ stands for the equivalent noise spectral density on gate $G_i$ at 1 Hz. 
We fit the experimental data with the two measurable electric susceptibilities, and extract two noise spectral densities $S_{T_4}^{1/2} = 2\pm 1\rm \mu V/\sqrt{Hz}$ and $S_{T_3}^{1/2} = 6\pm 1\rm \mu V/\sqrt{Hz}$. The fit is shown as a solid orange line in Fig.~\ref{fig4} and captures reasonably well the measured $T_2^E$ (orange points).

As evidenced by Eq.~\eqref{eq:t2echo}, the protection from \textit{dynamical} uncorrelated fluctuations can only be reached when all LSES are minimized, since the coherence time involves $\sum _{G_i}\beta_\parallel(G_i)^2$. On the other hand, the thermal susceptibility originates from \textit{static} sensitivity to electric fields, as underscored by Eq.~\eqref{eq:freqshift_model}. It is therefore minimized whenever the algebraic sum of LSES with different signs cancels out. 

This distinction accounts for the different field angles at which thermal and noise sweet spots are observed in this work. That said, operating spin qubits near an electrical sweet spot requires minimizing all LSES contributions. Consequently, this optimization is inherently beneficial for LTS, meaning that resilience to charge noise goes hand-in-hand with a relative resilience to heating effects, even when the two sweet spots do not perfectly coincide.

The ideal sweet spot condition would be the one where all LSES vanish, ensuring protection against both charge noise and local heating. Recent work has demonstrated that the anisotropy of spin-electric coupling in hole spin qubits depends on their electrostatic confinement, which can be individually tuned\cite{bassi_optimal_2026}. Tuning a large ensemble of qubits to an ideal sweet spot seems challenging though. A more feasible approach would be to tune the qubit ensemble to a “quasi-sweet-spot” where only the dominant LSES contributions are simultaneously suppressed. This would generally yield longer spin coherence alongside a significantly reduced thermal susceptibility.

\section{Conclusion}

We have investigated the thermal susceptibility of a hole spin qubit by measuring the temperature dependence of its Larmor frequency under different magnetic field orientations. The observed shifts correlate with the longitudinal spin-electric susceptibility, unambiguously pointing towards an electric origin. We introduced a microscopic model based on thermally activated dipoles in the oxide layer and supported it with realistic device-level simulations, which can faithfully reproduce our experimental observations. The accurate modeling of the local electric field allows us to extract physically meaningful values for the dipole moments and density. 
We argue the role of thermally activated dipoles evidenced here parallels the impact of two-level systems in other solid-state platforms, such as superconducting circuits \cite{muller_towards_2019,thorbeck_two-level-system_2023,spiecker_two-level_2023,odeh_non-markovian_2025} and mechanical resonators \cite{ bachtold_mesoscopic_2022,metzger_tls-induced_2025}, where microscopic defects are known to hamper coherence and gate fidelity.

In the case of semiconductor spin qubits investigated here, the Larmor frequency shift induced by local heating leads to computational errors. 
While coherent errors can be corrected via pulse engineering \cite{takeda_optimized_2018}, these techniques leave the qubit sensitive to incoherent errors, arising from stochastic or fast thermal fluctuations. 
By tuning the electrical susceptibility, we demonstrate the existence of a thermal sweet spot, at which the Larmor frequency becomes temperature independent.
At the sweet spot, the thermal error channel is suppressed at its origin, thus eliminating the need for active correction and making this approach particularly compelling to explore.

 While the sweet spot orientation is intrinsically device- and qubit-dependent, the different LSES have recently been shown to be largely tunable by means of the gate electrodes determining the electrostatic confinement.\cite{bassi_optimal_2026}
This offers prospects to finely tune the sensitivity to electric fields, and engineer in situ the system to align sweetspots accross a qubit array. More broadly, it suggests that an optimal sweet spot, where dominant susceptibilities are simultaneously minimized, could be approached through electrostatic engineering, rendering the hole spin intrinsically insensitive to both dynamical charge noise and thermally induced frequency shifts.

\section*{Data availability}
The data that support the findings of this article are openly available at \url{https://doi.org/10.5281/zenodo.17120293}.

\section*{Acknowledgments}
This research has been supported by the European Union’s Horizon 2020 research and innovation programm under grant agreement nos. 951852 (QLSI project), nos. 101174557 (QLSI2),  810504 (ERC project QuCube), and by the French National Research Agency (ANR) through the PEPR PRESQUILE (ANR-22-PETQ-0003). V.C. acknowledge support from the Program QuantForm-UGA n° ANR-21-CMAQ-0003 France 2030 and by the LabEx LANEF n° ANR-10-LABX-51-01. V.C. and B.Br. acknowledge support from ANR-23-CPJ1-0033-01.

\section*{Author contribution}
V.C. performed the measurements and analyzed the data with input from B.Br.. G.B. and B.M. deployed the numerical model. B.Be. and H.N. were responsible for the fabrication of the device. All authors participated in the discussions and development of the models. V.C., B.Br. and B.M. co-wrote the manuscript with inputs from all the authors.

\appendix

\section{g-matrix formalism}\label{app:fit}
We fit the g-factors and LSESs within the g-matrix formalism\cite{piot_single_2022,bassi_optimal_2026}:
\begin{align}
    g &= \sqrt{\vec{b}\cdot G\vec{b}}\\
    \beta_\parallel&= \frac{\mu_B B}{2hg} \vec{b}\cdot G'\vec{b}
\end{align}
where $\vec{b}$ is a unit vector along the magnetic field direction, $\mu_B$ is the Bohr magneton, $h$ is Planck's constant, $G$ and $G'$ are $2\times2$ matrices used as fit parameters (as we sweep the magnetic field in a single plane).

\section{Inhomogeneous Dephasing Times}\label{app:T2}

\begin{figure}
    \centering
    \includegraphics{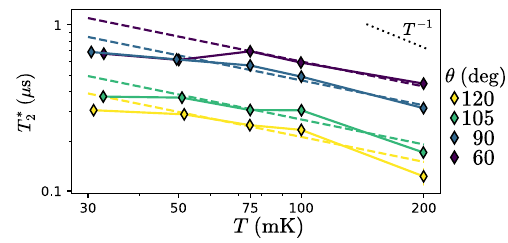}
    \caption{Temperature dependence of the coherence times for each orientation of the magnetic field on a log-log scale. The dashed line corresponds to fits to a $T^{-1/2}$ model. The black dotted line represents a $T^{-1}$ scaling for comparison.}
    \label{figT2}
\end{figure}

We extract the inhomogeneous dephasing time $T_2^*$ from the Ramsey measurements and plot its temperature dependence in Fig.~\ref{figT2} for all magnetic field orientations. First, we observe that the dephasing time increases with decreasing $|\beta_\parallel(T_4)|$, as expected for charge noise-limited coherence.

We expect the noise spectrum that arises from the fluctuations of a uniform distribution of two level systems to scale as \cite{connors_low-frequency_2019,burstein_semiconductor_2012} $S \propto k_BT/f$. Assuming furthermore that the noise is uncorrelated across the device, we can write \cite{ithier_decoherence_2005}:
\begin{equation}
    \frac{1}{T_2^*} \propto \sqrt{\sum_{G_i}\beta_{\parallel}(G_i)^2 S_{G_i}} \propto \sqrt{T\sum_{G_i}\beta_{\parallel}(G_i)^2}. \label{eq:scaling_t2}
\end{equation}
We plot in Fig.~\ref{figT2} the scaling in $ T^{-1/2}$ as dashed lines, showing good agreement with the data.

\section{Overhauser field}\label{app:hyperfin}
While the explanation by electric dipoles looks appealing, other contributions could also come into play. For instance, the present device is made out of natural silicon constituted of $4.9~\%$ of $^{29}\rm Si$. This isotope carries a nuclear spin and has a gyromagnetic factor $\gamma = -8.46~\rm MHz/T$, leading to a population of up-spin states $P_{nuc} = \tanh \left ( \frac{\gamma B}{2k_B T} \right )$ where $B$ is the external magnetic field and $T$ is the temperature. At $B=500~\rm mT$, the population in the excited state changes by $0.28\%$ between $30$ and $200~\rm mK$, which could slightly renormalize the Larmor frequency through hyperfine interaction. 

The hyperfine constant $A$, which quantifies the shift in the Larmor frequency due to nuclear spins, is on the order of $100~\rm neV$ for a hole in silicon \cite{assali_hyperfine_2011,philippopoulos_first-principles_2020}. The hyperfine shift is proportional to the overlap of the charge wavefunction with nuclear spins and, therefore, strongly depends on the confinement of the charge, and consequently on the geometry of the device, etc. However, we can compute an order of magnitude of the expected frequency shift as $ \delta f_L(T) = \frac{2 A }{h}\delta P_{nuc}(T)$. This yields a frequency shift $\delta f_L \sim 140~\rm kHz$  between $30$ and $200~\rm mK$. This estimate is too small to fully explain our data, which reaches up to $1.2~\rm MHz$. However, it is not negligible, especially for magnetic field orientations where the thermal susceptibility is mitigated.

\section{Simulated LSES for gate $T_3$}\label{app:LSEST3}

\begin{figure}
    \centering
    \includegraphics[scale =0.8]{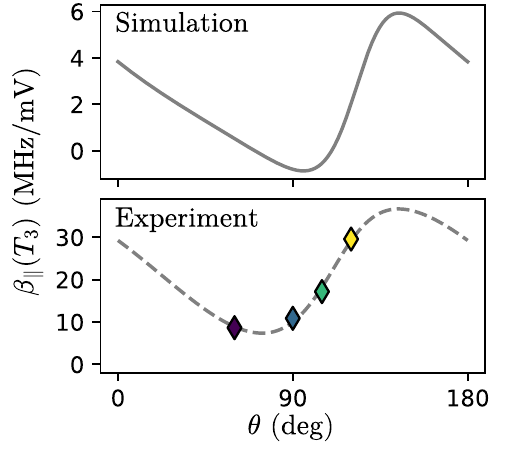}
    \caption{Top panel: Simulated $\beta_\parallel(T_3)$ as a function of the magnetic field orientation. The simulated device is a larger version of that shown in Fig.~\ref{fig2}a that includes all four side gates $T_3$, $B_3$, $T_5$, $B_5$. Bottom panel: measured $\beta_\parallel(T_3)$ and fit.}
    \label{fig_supp_lsest3}
\end{figure}

In this Appendix we compute $\beta_\parallel$ for gate $T_3$, and compare it to the experimental data shown in Fig.~\ref{fig1}b. The device simulated in the main text only includes half of the gates next to $T_4$ and $B_4$ (see Fig.~\ref{fig2}a), which are in fact the same gate on the left and right due to periodic boundary conditions. This reduced device is well suited to the efficient calculation of the potential of  dipoles, since only those close to the central gates have a sizable effect on the QD. However, it does not allow to compute the LSES for any of the side gates. 

To evaluate $\beta_\parallel(T_3)$ we have simulated a slightly larger device including all four side gates $T_3$, $B_3$, $T_5$, $B_5$ so that they can be independently biased. The results are shown in Fig.~\ref{fig_supp_lsest3}, where we recover the same behavior as the experimental data shown in Fig.~\ref{fig1}b despite some quantitative differences. In particular there is a change of sign in the simulations that is not observed experimentally (as for $\beta_\parallel(T_4)$). In addition, the calculated LSES is significantly smaller than the experimental one. These differences, together with those in $\beta_\parallel(T_4)$, explain the quantitative discrepancies between experiment and simulation in Fig. \ref{fig3}c.

\section{Robustness of the dipole model beyond the homogeneous dipole moment approximation} \label{app:model_hom}

\begin{figure}[t!]
    \centering
    \includegraphics[width=0.9\columnwidth]{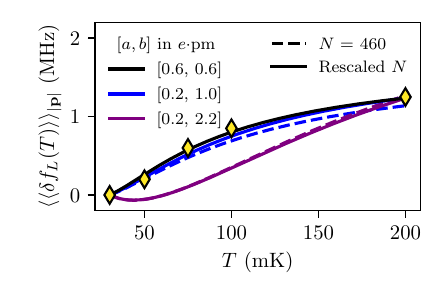}
    \caption{Temperature dependence of the Larmor frequency drift averaged over an uniform distribution of $|\vec{p}|$ within a $[a,b]$ window for the magnetic field orientation $\theta = 120$ deg in experiments and simulations). Dashed lines show the data for $N=460$. Solid lines illustrate the trends with a dipole density rescaled by a factor 1.09 and 0.97 for $[a,b]=[0.2,1.0]$ and $[0.2,2.2]$ $ \rm e\cdot$pm, which correspond to $\langle|\vec{p}| \rangle = 0.6$ and $\langle|\vec{p}| \rangle = 1.2~e\cdot$pm respectively. The diamonds are the experimental data. In black we provide the reference case of homogeneous $|\vec{p}|$ across all dipoles.}
    \label{fig:app_qd}
\end{figure}

The implemented dipole model imposes the same moment $|\vec{p}|$ for all dipoles, which may seem a strong and unrealistic simplifying assumption. In this Appendix we demonstrate that the experimental data remains compatible with a dipole model accounting for scattered dipole moments as long as $T_0$ (and thus $\langle|\vec{p}| \rangle = 0.6~e\cdot$pm) is conserved. We therefore cannot retrieve information on the distribution of $|\vec{p}|$ in the experiment, but we observe that a dipole distribution with $\langle|\vec{p}| \rangle \approx 1.2~e\cdot$pm substantially differs from experiments, which strengthens the validity of the $\langle|\vec{p}| \rangle \approx 0.6~e\cdot$pm estimate.

We can consider that the individual dipole moments $|\vec{p}_i|$ follow a probability distribution $P(|\vec{p}|)$. Given an expression for $P(|\vec{p}|)$ we can compute the Larmor frequency shift of an individual dipole $i$ averaged over $|\vec{p}|$,

\begin{equation}
    \langle \delta f_{L,i} (T) \rangle_{|\vec{p}|} = \int P(|\vec{p}|) \delta f_{L,i} (T,|\vec{p}|)   d(|\vec{p}|) .
    \label{eq:int_d}
\end{equation}
$|\vec{p}|$ is by design a positive quantity, and its distribution is \textit{a priori} little constrained. Here we consider for illustration purposes a uniform distribution within the $[a,b]$ range, 
\begin{equation}
P(|\vec{p}|) = 
\begin{cases} 
\frac{1}{b-a} & \text{if } a \leq |\vec{p}| \leq b,  \\
0 & \text{otherwise.}
\end{cases}
\label{eq:uniform_d}
\end{equation}
Unfortunately, the integral in Eq.~\eqref{eq:int_d} does not have analytical solution even for this simple ansatz, but we can still compute it numerically for different values of $a$ and $b$, and analyze the impact of the inhomogeneous dipole moments on the final results. We define $\langle \langle \delta f_L (T) \rangle \rangle _{|\vec{p}|}$ as the Larmor frequency shift $\delta f_L^{(N)}(T)$ averaged over the ensemble of dipole baths, where for each dipole bath $\delta f_L^{(N)}(T)= \sum_{i=1}^N \delta f_{L,i}(T)$. 

In Fig.~\ref{fig:app_qd} we show the obtained results. It is important to note that $|\vec{p}|$ has a direct impact on the dipoles activation temperature, as $T_0 \approx \langle |\vec{p}_i \cdot \vec{E}_i| \rangle/k_B$ in absence of intrinsic detuning. Obtaining the right temperature dependence of $\langle\langle \delta f_L (T) \rangle \rangle_{|\vec{p}|}$ therefore requires a symmetric distribution of $|\vec{p}_i|$ centered at  $\langle |\vec{p}| \rangle=\frac{a+b}{2} = 0.6~e\cdot$pm. The  blue curve illustrates the results for $a=0.2$ and $b=1.0~e\cdot$pm. In this case, the obtained results are barely distinguishable from the experimental data and from the homogeneous dipole approximation. As a consequence, we cannot draw conclusions on the distribution of $|\vec{p}|$ in the experiment. 

Alternatively, we may estimate the uncertainty associated with the provided average dipole moment. Dipole distributions with $\langle |\vec{p}| \rangle > 0.6~e\cdot$pm are expected to display higher $T_0$. However, given the limited resolution of the experimental data, we may assess to what extent we can distinguish a dipole bath with $\langle |\vec{p}| \rangle = 0.6~e\cdot$pm from a bath with $\langle |\vec{p}| \rangle > 0.6~e\cdot$pm and a rescaled dipole density. To do so, we consider a $P(|\vec{p}|)$ distribution with $\frac{a+b}{2} > 0.6~e\cdot$pm, and we compare the temperature dependence of $\langle \langle \delta f_L (T) \rangle \rangle _{|\vec{p}|}$ to the experimental data. The results for $\langle |\vec{p}| \rangle = 1.2$ $e\cdot$pm are shown in purple in Fig.~\ref{fig:app_qd}, and highlights that a $\times 2$ increase in $\langle |\vec{p}| \rangle$ already gives rise to noticeable disagreements with the experimental data. 

Therefore, we can conclude that the proposed dipole model is valid beyond the homogeneous dipole moment approximation. A dipole ensemble with scattered dipole moments remains compatible with the experimental evidence as long as $\langle |\vec{p}| \rangle \approx 0.6~e\cdot$pm. Moreover, the actual shape of the distribution does not significantly alter the temperature dependence of the Larmor frequency shift within the experimental range.

\section{Robustness of the dipole model in presence of finite intrinsic detuning} \label{app:model}

All the numerical results in the main text neglect the effect of the intrinsic detuning $\varepsilon_{0,i}$. In reality, however, $\varepsilon_{0,i}$ are most likely scattered, and we may find dipoles for which $\varepsilon_{0,i}$ is comparable or even larger than $\mathbf{p}_i \cdot \mathbf{E}_i$.  For the relevant dipoles, $\mathbf{p}\cdot \mathbf{E}$ must be the dominant term of the energy. Indeed, dipoles need to be systematically aligned with $\mathbf{E}$ when polarized to achieve the structural order that generates a temperature-dependent net electric field (and therefore a net $\delta f_L$). According to Eq.~\eqref{eq:temp_dep}, dipoles with $|\varepsilon_{0,i}| \gg |\mathbf{p}_i\cdot \mathbf{E}_i|$ would display higher activation temperatures, and would not even yield to a temperature-dependent $\langle \delta f_L(T) \rangle$. It is not obvious, though, whether in the case $|\varepsilon_{0,i}| \sim |\mathbf{p}_i\cdot \mathbf{E}_i|$ (where $\langle |\varepsilon_i| \rangle$ is weakly affected), the complex interplay between the two terms can break the conclusions drawn in the main text. In this Appendix, we demonstrate that $\varepsilon_{0,i}$ has almost no impact on the temperature dependence and essentially renormalizes the dipole density $\rho$ as long as $T_0 = \langle |\varepsilon_i| \rangle/k_B \approx 70$ mK is preserved. 

For a given dipole $i$ and a certain ansatz for the probability distribution $P(\varepsilon_{0})$, we can integrate Eq.~\eqref{eq:temp_dep} to obtain the thermal susceptibility averaged over all possible $\varepsilon_{0}$,

\begin{equation}
     \langle\delta f_{L,i} (T) \rangle_{\varepsilon_0} = \int P(\varepsilon_0) \delta f_{L,i} (T,\varepsilon_0) d(\varepsilon_0) .
    \label{eq:int_AE0}
\end{equation}
Some constraints apply to $P(\varepsilon_0)$. If we assume the dipole orientation to be completely random in absence of electric field, $P(\varepsilon_0)$ must be a symmetric distribution centered at zero. Let us assume the simplest candidate: a uniform distribution within an energy range $\pm a$,

\begin{equation}
P(\varepsilon_0) = 
\begin{cases} 
\frac{1}{2a} & \text{if } -a \leq \varepsilon_0 \leq a, \\
0 & \text{otherwise.}
\end{cases}
\label{eq:uniform}
\end{equation}
With this ansatz the integral in Eq.~\eqref{eq:int_AE0} is analytical,

\begin{equation}\label{eq:result_int}
    \langle   \delta f_{L,i} (T)  \rangle_{\varepsilon_0} =  -\Delta f_{L,i} \frac{k_B T}{2a}  \ln \left( \frac{\cosh\left(\frac{a - \vec{p} \cdot \vec{E}_i}{k_B T}\right)}{\cosh\left(\frac{a + \vec{p} \cdot \vec{E}_i}{ k_B T}\right)} \right),
\end{equation}
which allows for a straightforward assessment of the effects of the interplay between $\varepsilon_0$ and $\vec{p}\cdot\vec{E}$. As in Appendix \ref{app:model_hom}, we define $\langle \langle \delta f_{L} (T) \rangle \rangle_{\varepsilon_0}$ as the Larmor frequency shift averaged over the dipole baths,  with each individual dipole contribution following Eq.~\eqref{eq:result_int}.

\begin{figure}
    \centering
    \includegraphics[width=0.9\columnwidth]{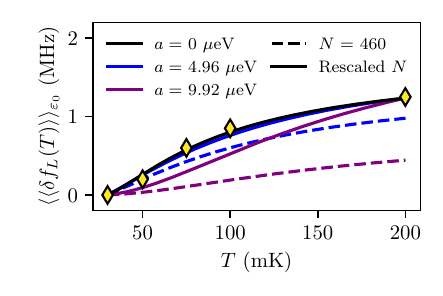}
    \caption{Temperature dependence of the Larmor frequency drift averaged over a uniform distribution of $\varepsilon_0$ within a $\pm a$ window for the magnetic field orientation $\theta = 120$ deg in experiments and simulations. Dashed lines show the data for $|\vec{p}| = 0.6$ $e \cdot$pm and $\rho = 3.54 \times 10^{18}$~cm$^{-3}$. Solid lines illustrate the trends with a dipole density rescaled by a factor of 1.26 and 2.78 for $a=\langle| \vec{p}\cdot\vec{E}_i | \rangle = $ 4.96 $\mu$eV and $a=2\langle | \vec{p}\cdot\vec{E}_i | \rangle = $ 9.92 $\mu$eV respectively. In black we provide the case of $\varepsilon_0 = 0$. The diamonds are the experimental data.}
    \label{fig:app_AE0}
\end{figure}

In Fig.~\ref{fig:app_AE0} we show $\langle \langle \delta f_L (T) \rangle \rangle_{\varepsilon_0}$ for the $\theta$ with the largest Larmor frequency shift (yellow data in Fig.~3c). We have chosen parametric values for $a$ that illustrate both $T_0 = \langle | \varepsilon_i | \rangle/k_B \approx 70$ mK ($a = \langle |\mathbf{p}\cdot \mathbf{E}_i | \rangle = 4.96$ $\mu$eV, blue curve in Fig.~\ref{fig:app_AE0}) and $T_0 = \langle |\varepsilon_i |\rangle/k_B > 70$ mK ($a = 2\langle |\mathbf{p}\cdot \mathbf{E}_i | \rangle = 9.92$ $\mu$eV, purple curve in Fig.~\ref{fig:app_AE0}). The dashed lines are the results for $\rho = 3.54 \times 10^{18}$~cm$^{-3}$, while the solid lines are the results with $\rho$ rescaled to match the reference $\delta f_L$ at $\varepsilon_0=0$ and $T=200$ mK (black line).

We observe that averaging over $\varepsilon_0$ reduces the Larmor frequency shift. However, for $a=4.96$ $\mu$eV the experimental temperature dependence is recovered when rescaling the density by a factor of 1.26. The experimental data does not provide enough resolution to make an unequivocal distinction between this case and $\varepsilon_0 = 0$. Indeed, with $\langle |\varepsilon_{0,i} |\rangle = a/2$, $\langle |\varepsilon _i|\rangle$ is weakly affected even in the limit of $a=\langle |\mathbf{p}\cdot \mathbf{E}_i | \rangle$, as most dipoles still fulfill $|\varepsilon_{0,i}| < |\mathbf{p}\cdot \mathbf{E}_i|$. For $a=2\langle |\mathbf{p}\cdot \mathbf{E}_i | \rangle$, $\langle |\varepsilon _0|\rangle=|\mathbf{p}\cdot \mathbf{E}_i|$ and there are more dipoles displaying a higher activation temperature. As a consequence, $\langle |\varepsilon _i|\rangle$ is indeed larger, and the average $T_0$ increases. Even when rescaling to match the Larmor drift at $T=200$ mK we observe significant differences with the experimental data. 

These results highlight that a dipole bath with scattered, non-zero $\varepsilon_i$'s is still compatible with the experimental data as long as $\mathbf{p}\cdot \mathbf{E}_i$ remains the dominant term. This justifies neglecting the intrinsic detuning to estimate $|\vec{p}|$ and $\rho$ in the main text. Moreover, although $\varepsilon_0$ may indeed be finite, it must not be larger than a few $\mu$eV. This imposes a strong constraint to the origin of these dipoles. 

Alternatively, we have considered more complex expressions for $P(\varepsilon_0)$ by introducing a correlation between $|\vec{p}|$ and $\varepsilon_0$, assuming that larger intrinsic detunings may yield to larger dipole moments. With this approach, we have been able to scatter $\varepsilon_0$ (and $|\vec{p}|$) over orders of magnitude and still recover the right LTS in some cases. However, results are strongly dependent on the chosen correlation, on which we have no experimental information. 

\section{Quantification of the uncertainty in $|\vec{p}|$ due to specificities of individual dipole baths} \label{app:d}

\begin{figure}
    \centering
    \includegraphics[width=0.9\columnwidth]{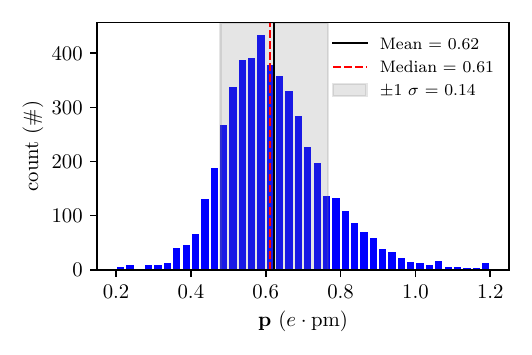}
    \caption{Histogram of the dipole moments resulting from the individual fits that ensure $T_0=70$ mK for all generated dipole distributions in Fig.~\ref{fig3}a.}
    \label{fig:app_diff_d}
\end{figure}

The estimated dipole moment given in the main text has been obtained by fitting the average Larmor frequency shift $\langle \delta f_L \rangle$ to the experimental results. There are, however, clear fingerprints of variability in the $\delta f_L$ of individual dipole distributions, and there is no guarantee that the experimental device is actually representative of the maximum likelihood in Fig.~\ref{fig3}a. Moreover, the Larmor frequency shifts in Fig.~\ref{fig3}a have been computed with the same $|\vec{p}|$ for each dipole distribution. The reported $|\vec{p}|$ indeed guarantees that the activation temperature $T_0$ matches the experiments on average, but not necessarily for all distributions. In this Appendix, we independently fit $|\vec{p}|$ for each dipole distribution to ensure $T_0=70$~mK, and collect statistics on the dipole moment. This gives an estimate of the uncertainty on $|\vec{p}|=0.6$~$e\cdot$pm that arises from the unknown location of the experimental device in the distribution of simulated dipole baths. 

In Fig.~\ref{fig:app_diff_d} we provide the histogram of calculated $|\vec{p}|$'s for $\theta=120$ deg. Although values get scattered around $|\vec{p}|=0.6$ $e\cdot$pm, the spread of the distribution remains fairly small (standard deviation $\sigma \sim 0.1$ $e\cdot$pm). These results show that the order of magnitude of $|\vec{p}|$ is hardly affected by the specificities of the dipole bath in the experimental device.

\section{Simulation of randomized benchmarking with thermal errors} \label{app:RB}
In this appendix, we quantify the impact of thermally induced frequency shifts on single-qubit gate fidelity using a simple randomized benchmarking (RB) model\cite{knill_randomized_2008,lawrie_simultaneous_2023,muhonen_quantifying_2015}.

We consider a two-level system driven resonantly at frequency $f_0$ with Rabi frequency $f_R$. In the presence of a static detuning $\delta f$ (e.g., induced by a thermally shifted Larmor frequency), the Hamiltonian in the rotating frame reads

\begin{equation}
H = \frac{h}{2} \left( \delta f \, \sigma_z + f_R \, \sigma_x \right).
\end{equation}

The drive no longer generates a pure rotation about the $x$ axis, but instead about an axis in the $x$–$z$ plane tilted by:

\begin{equation}
\phi = \arctan\left( \frac{\delta f}{f_R} \right).
\end{equation}

In addition, the net Rabi frequency is increased, such that a nominal $\pi$ pulse of duration $\tau = 1/(2 f_R)$ gives rise to an over-rotation

\begin{equation}
\theta = \pi \sqrt{1 + \left( \frac{\delta f}{f_R} \right)^2 }.
\end{equation}

Both effects contribute to coherent gate errors.

We model primitive $X$ and $Y$ rotations as imperfect unitaries generated by the above Hamiltonian with detuning $\delta f$. The corresponding unitary is, for a gate of duration $\tau = 1/(2f_R)$,

\begin{equation}
U(f_R,\delta f) = \exp\left[-i \pi  \left(\sigma_{x,y} + \frac{\delta f}{f_R} \sigma_z \right)/2 \right].
\end{equation}

To estimate the impact on experimentally relevant sequences, we simulate standard single-qubit Clifford randomized benchmarking. Sequences of random Clifford gates are generated, each decomposed into primitive $X$ and $Y$ $\pi/2$ and $\pi$ pulses. Each primitive gate is replaced by its detuned version described above. A recovery gate is computed from non-detuned pulses and applied subsequently. The recovery probability is averaged over many random sequences, and the decay parameter is extracted in the usual manner to obtain an effective average gate infidelity.

\begin{figure}
    \centering
    \includegraphics[width=1\columnwidth]{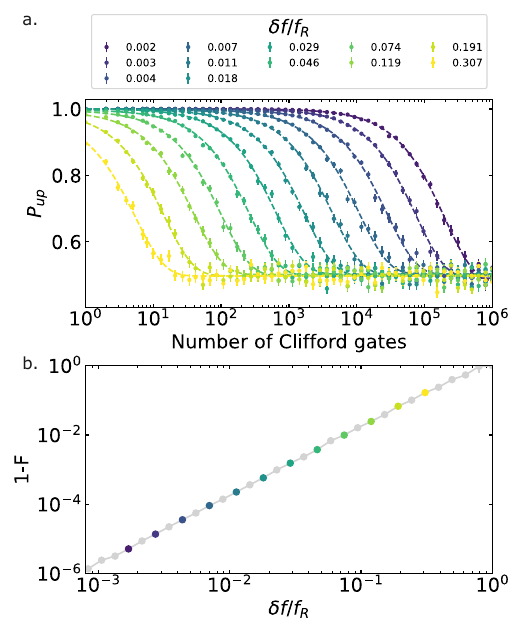}
    \caption{\textbf{Randomized benchmarking. a.} Recovery probability $P_{up}$ as a function of the number of Clifford gates for various ratios $\delta f/f_R$. \textbf{b.} Gate infidelity as a function of the ratio $\delta f/f_R$. Colors correspond to the traces shown in panel \textbf{a.}, while grey markers indicate additional simulated points, that are omitted in \textbf{a.} for clarity.}
    \label{fig:app_benchmarking}
\end{figure}

Figure~\ref{fig:app_benchmarking}a shows the recovery probability $P_{up}$ obtained from the randomized benchmarking simulations for various ratios $\delta f/f_R$ between the Larmor frequency shift and the Rabi frequency. The extracted gate infidelity $1-F$ is plotted in Fig.~\ref{fig:app_benchmarking}b as a function of $\delta f/f_R$. The infidelity scales as $(\frac{\delta f}{f_R})^2$ as expected from analytical analysis of single gate errors in Eq.(H4).
The simulations indicate that achieving single-qubit gate fidelities above $99.9\%$ requires maintaining $\delta f/f_R \lesssim 3\%$. For comparison, Ref.~\cite{undseth_hotter_2023} reports thermally induced frequency shifts corresponding to ratios of a few tens of percent of the Rabi frequency. These values therefore lie in a regime where thermally induced detuning can already produce gate errors approaching the $10^{-2}$ level and above.
The analysis above assumes a quasi-static detuning during each gate. In this case, the induced error is coherent and, in principle, correctable through calibration or adaptive control. However, if the thermal detuning fluctuates on timescales comparable to or shorter than gate durations, the resulting errors acquire an incoherent component that cannot be fully mitigated by pulse-level correction. This distinction further motivates reducing the underlying longitudinal thermal susceptibility at the device level.

\end{document}